\documentclass[aps,prl,reprint,superscriptaddress]{revtex4-2}
\renewcommand{\selectlanguage}[1]{} 

\usepackage[utf8]{inputenc}
\usepackage[T1]{fontenc}
\usepackage{lmodern}
\usepackage{amsmath,amsfonts,amssymb}
\usepackage{graphicx}

\usepackage{stackengine}
\stackMath

\usepackage{siunitx}

\usepackage{physics2} 
\usephysicsmodule{qtext.legacy, braket, diagmat, xmat, ab}
\usepackage{soul}

\usepackage[capitalise]{cleveref} 
\Crefname{equation}{Eq.}{Eqs.} 

\usepackage{empheq} 
\usepackage{dsfont} 

\newcommand{\jj}{\mathrm{j}} 
\newcommand{\ee}{\mathrm{e}} 

\renewcommand{\bold}[1]{\boldsymbol{\mathbf{#1}}}  
\newcommand{\diff}{\mathop{}\!\mathrm{d}}
\usepackage{amsthm}

\usepackage{tikz}
\newcommand*\circled[1]{\tikz[baseline=(char.base)]{
            \node[shape=circle,draw,inner sep=1pt] (char) {#1};}}

\DeclareRobustCommand{\varlambda}{\text{\usefont{OML}{txmi}{m}{it}\symbol{"15}}}

\begin{document}


\title{Multi-Objective Tweezers in Scattering Media}


\author{Tristan Nerson}
\email{tristan.nerson@epfl.ch}
\affiliation{Laboratory of Wave Engineering, School of Electrical Engineering, EPFL, Lausanne, Switzerland}

\author{Jakob Hüpfl}
\affiliation{Institute for Theoretical Physics, Vienna University of Technology (TU Wien), A-1040 Vienna, Austria}

\author{Clément Ferise}
\affiliation{Laboratory of Wave Engineering, School of Electrical Engineering, EPFL, Lausanne, Switzerland}

\author{David Globosits}
\affiliation{Institute for Theoretical Physics, Vienna University of Technology (TU Wien), A-1040 Vienna, Austria}

\author{\mbox{Marlene Hudler}}
\affiliation{Institute for Theoretical Physics, Vienna University of Technology (TU Wien), A-1040 Vienna, Austria}

\author{Matthieu Malléjac}
\affiliation{Univ. Bordeaux, CNRS, Bordeaux INP, I2M, UMR 5295, F-33400, Talence, France}
\affiliation{Arts et Metiers Institute of Technology, CNRS, Bordeaux INP, I2M, UMR 5295, F-33400 Talence, France}

\author{Stefan Rotter}
\affiliation{Institute for Theoretical Physics, Vienna University of Technology (TU Wien), A-1040 Vienna, Austria}

\author{Romain Fleury}
\email{romain.fleury@epfl.ch}
\affiliation{Laboratory of Wave Engineering, School of Electrical Engineering, EPFL, Lausanne, Switzerland}


\date{\today}

\begin{abstract}
Radiation forces and torques enable the manipulation of objects with acoustic and electromagnetic waves. Yet, harnessing them in complex scattering media remains a formidable challenge, especially when multiple objects must be controlled under competing objectives. Here, we demonstrate that sound or light can be shaped to tailor momentum transfer to multiple objects simultaneously in a complex scattering medium. For a single object, our theory yields the maximal achievable force or torque; for multiple objects, it produces Pareto-optimal actuation and exact bounds on the simultaneous realization of incompatible objectives. This opens new applications for wave tweezers, enabling selective and precise manipulation of objects within complex media, ranging from the handling of cells, organoids, or microrobots, to targeted drug delivery in biological media.
\end{abstract}


\maketitle


When propagating through complex media, waves -- acoustic or electromagnetic, among others -- undergo repeated scattering events due to inhomogeneities, which challenges our ability to control and predict their propagation. Early work in acoustics showed that wavefronts can be time-reversed to refocus through complex media, revealing that multiple scattering does not necessarily lead to information loss \cite{fink_time-reversed_2000,horodynski2021invariance}. Subsequent optical experiments demonstrated that tailored wavefronts can focus light through opaque, strongly scattering media, marking the experimental inception of the field of wavefront shaping in optics \cite{vellekoop_focusing_2007,mosk_controlling_2012}, and further accelerated by the measurement of the optical transmission matrix \cite{popoffMeasuringTransmissionMatrix2010}. A key concept in modern wavefront shaping is the use of the Generalized Wigner-Smith (GWS) matrix, defined from the scattering matrix $\bold{S}$ as
\begin{equation} \label{eq:GWS}
    \bold{Q}_\alpha = \jj \bold{S}^\dagger \partial_\alpha{\bold{S}},
\end{equation}
where $\alpha$ denotes a given parameter of the scattering region, and $\dagger$ denotes the conjugate transpose (assuming the $+\jj \omega t$ time-harmonic convention). Originally introduced in quantum scattering \cite{smithLifetimeMatrixCollision1960, froissart_spatial_1963,brouwer_quantum_1997}, $\bold{Q}_\alpha$ has become an essential tool to construct incident states that extremize parametric responses \cite{ambichl_focusing_2017,horodynski_optimal_2020}, and has enabled a broad range of wave-control applications \cite{bouchetMaximumInformationStates2021a,matthes_learning_2021,delhougneCoherentWaveControl2021,horodynski_anti-reflection_2022,hupflOptimalCoolingMultiple2023a,solOptimalBlindFocusing2024,butaitePhotonefficientOpticalTweezers2024,globosits2024pseudounitary,goicoecheaDetectingFocusingNonlinear2025,byrnesPerturbingScatteringResonances2025,bliokhDynamicGeometricShifts2025}.

Beyond controlling the spatial distribution of wave amplitude or phase, many applications rely on the mechanical action that waves exert on objects. Contactless manipulation of microscopic objects is vital across biology, medicine, and materials science \cite{dholakia2011shaping}, but many envisioned uses -- selective handling of cells and organoids, targeted drug delivery, elaborate microrobot actuation, automated biofabrication, and mechanobiology -- remain constrained by the need for high precision in complex, inhomogeneous media. In acoustics, this precision has not been achieved in strongly scattering biological tissues, and existing in vivo demonstrations largely rely on focal targeting rather than programmable wave-momentum transfer \cite{ghanemNoninvasiveAcousticManipulation2020,loTornadoinspiredAcousticVortex2021a,yangInvivoProgrammableAcoustic2023}. High-fidelity operation typically remains tied to small, near-spherical, subwavelength objects, and concurrent manipulation in complex scattering environments is not yet broadly available \cite{baudoin2019folding,baudoin2020spatially,hirayama2022high,shen2024acousto}. In optics, holographic tweezers scale well in transparent media \cite{curtis2002dynamic,leach20043d,holman2026trapping}, and wavefront shaping has demonstrated trapping through scattering by iteratively reforming foci \cite{vcivzmar2010situ,peng2019real}. Yet refocusing alone does not provide a general, exact, one-shot route to simultaneously tailor the momentum of multiple scattering objects. Prior work established that, for a single object undergoing a small translation or rotation, extremal GWS states can maximize the corresponding radiation force or torque in a complex environment \cite{ambichl_focusing_2017,horodynski_optimal_2020,horodynskiTractorBeamsOptimal2023a,orazbayevWavemomentumShapingMoving2024}. What has been lacking, however, is a general theoretical framework that unifies these ideas for arbitrary acoustic or electromagnetic media and extends them to individual control of multiple objects of any shape, thus enabling a new range of applications leveraging optimal tweezers. 

\begin{figure}
    \centering
    \includegraphics{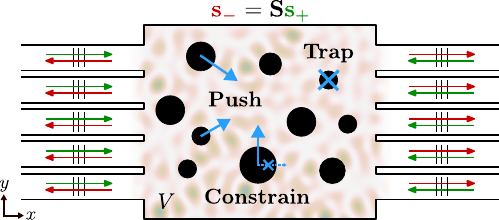}
    \caption{Concept of multi-objective tweezers in scattering media. A multiport disordered cavity (scattering region $V$) is driven through accessible input channels, with incident $\bold{s}_+$ and outgoing $\bold{s}_-$ complex modal amplitudes connected through the $\bold{S}$-matrix. By tailoring incident wavefronts, the interference inside the cavity is controlled so that radiation forces on embedded objects are simultaneously programmed: multiple targets are optimally pushed; one object is trapped at a prescribed location; one object’s motion is constrained to a single direction.}
    \label{fig:scattering_media}
\end{figure}

Here, we develop such a framework based on variational relations linking parametric changes of the medium to modifications of the far-field scattering amplitudes. This naturally leads to GWS matrices that encode individual forces and torques on multiple objects at once. Our multi-objective formulation establishes that trade-offs between competing mechanical targets are inherently unavoidable, but also highly relevant in applications. The simultaneous optimization of such incompatible objectives is fundamentally constrained by a theoretical bound, which quantifies the achievable performance limits. Within this framework, the eigenstates of the GWS provide Pareto-optimal solutions, enabling the coordinated manipulation of multiple objects. In particular, we demonstrate how tailored input wavefronts can be used to achieve coordinated and selective control over the motion of multiple objects. Examples include simultaneously optimizing pushes on several objects along prescribed directions, trapping objects at chosen locations, and imposing directional constraints, such as maximizing the force along the $y$-direction while suppressing motion along $x$, and keeping other objects still (\cref{fig:scattering_media}).

\paragraph{Radiation forces in inhomogeneous media---} 
We first set the stage by introducing time-averaged body force density expressions under harmonic excitation in lossless media. In acoustics \cite{karlsenAcousticForceDensity2016},
\begin{equation} \label{eq:force_density_ac}
\overline{\bold{f}}_\text{ac} = - \frac{1}{4} \ab( \ab|p_1|^2 \nabla \kappa_0 + \ab|\bold{v}_1|^2 \nabla\rho_0),
\end{equation}
where $\kappa_0$, $\rho_0$, $p_1$ and $\bold{v}_1$ are the compressibility, density, first order complex pressure and acoustic velocity fields, respectively. Although initially derived for inhomogeneous fluids, it generally applies when solid scatterers can be modeled as equivalent fluids, i.e., transverse waves are neglected. A rule of thumb is to consider solid inclusions of characteristic size $d$ in which $\omega \ll \pi c_s/ d$, where $c_s$ is the shear speed of sound. Analogously, in electromagnetism \cite{anghinoniFormulationsElectromagneticStress2022},
\begin{equation}\label{eq:force_density_em}
    \overline{\bold{f}}_\text{em} = - \frac{1}{4} \ab( |\bold{E}|^2 \nabla \varepsilon + |\bold{H}|^2  \nabla \mu),
\end{equation}
where $\varepsilon$, $\mu$, $\bold{E}$ and $\bold{H}$ are the permittivity, permeability, complex electric and magnetic vector fields, respectively. This is valid in linear, isotropic, non-dispersive and lossless inhomogeneous media with no free charges or currents. By exploiting the analogy between \cref{eq:force_density_ac,eq:force_density_em}, we formulate a general theory of radiation forces in scattering media from far-field quantities.

\paragraph{Variational relations---}
Following circuit theory nomenclature \cite{zemanian1963n,dickeGeneralMicrowaveCircuit1987}, we introduce some scattering variables called immittance voltage $u_m$ and current $i_m$, defined at a scattering port $m$ as $u_m=\sqrt 2\ab(s_m^{+}+s_m^{-})$ and $i_m=\sqrt2\ab(s_m^{+}-s_m^{-})$, where $s^\pm_m$ is the incident ($+$) and outgoing ($-$) complex amplitude of the wave, normalized such that $\ab|s_m^\pm|^2$ is equal to the incident/outgoing power at the corresponding port. In the linear scattering regime, the $\bold{S}$-matrix is defined as $\bold{s}_- = \bold{S} \bold{s}_+$, so the immittance vectors read
\begin{equation} \label{eq:immittance_def}
\mathbf{u} =\sqrt2\ab(\mathds{1}+\mathbf{S})\mathbf{s}_+, \quad \mathbf{i}=\sqrt2\ab(\mathds{1}-\mathbf{S})\mathbf{s}_+,
\end{equation}
where $\mathds{1}$ denotes the identity matrix. In practice, the $N$ ports correspond to the asymptotic modes that couple the scattering region to its environment. In the open cavity of \cref{fig:scattering_media}, they are the propagating waves in the monomode waveguides coupled to the cavity. The far-field information about the scattering medium of volume $V$ is completely described by the $\bold{S}$-matrix with respect to these ports (details in Supplemental Material \cite{supplemental}).

We are interested in the far-field signature when a parameter $\bold{\alpha}$ (e.g., the position of a target) in the system is slightly varied. In the Appendix A, we show that the variations of the immittance vectors at the ports ($\delta \bold{i}$ and $\delta\bold{u}$) can be linked to the local changes in the scattering medium in both acoustics and electromagnetism, i.e.,  
\begin{equation} \label{eq:main_varia}
    \frac{\bold{u}^\dagger \delta \bold{i} + \bold{i}^\dagger \delta \bold{u}}{\jj \omega} = \delta \bold{\alpha} \cdot \! \int_V \! \ab(\ab|\stackanchor{p_1}{\bold{E}}|^2 \nabla_{\scriptscriptstyle{\!\! \delta \bold{\alpha}}} \ab(\stackanchor{\kappa_0}{\varepsilon}) +\ab|\stackanchor{\bold{v}_1}{\bold{H}}|^2 \nabla_{\scriptscriptstyle{\!\! \delta \bold{\alpha}}} \ab(\stackanchor{\rho_0}{\mu}) ) \! \diff V \!\!.
\end{equation}
As we show below, these variations generate local radiation forces. We illustrate the argument with the density field $\rho_0$; the same construction applies to $\kappa_0,\epsilon$ and $\mu$. Consider a system built from $N_o$ arbitrarily-shaped objects of constant density $\rho_{0,i}$ in a background $\rho_0^{(b)}(\mathbf{r})$, such that
\begin{equation}
    \rho_0(\mathbf{r})=\rho_0^{(b)}(\mathbf{r})+\sum_{i=1}^{N_o}\chi_i(\bold{r})\ab(\rho_{0,i}-\rho_0^{(b)}(\mathbf{r})).
\end{equation}
Here, $\chi_i(\bold{r})$ represents the indicator function of object $i$, taking a value of 1 inside the object and 0 outside. When object $k$ undergoes a small displacement by a vector $\delta \bold{\alpha}=\delta \bold{r}$  its indicator function shifts accordingly as $\chi_k(\bold{r})\rightarrow\chi_k(\bold{r}-\delta \bold{r})$. The induced variation is supported only in a local region $V_p$ containing the displaced interface, so the $\nabla_{\! \delta \bold{\alpha}}$ terms in \cref{eq:main_varia} vanish outside $V_p$ and the integral may be restricted from $V$ to $V_p$. Within $V_p$, we assume that $\rho_0^{(b)}$ is constant, and thus obtain $\nabla \rho_0 = - \nabla_{\! \delta \bold{r}} \rho_0$ with $\nabla_{\! \delta \bold{r}} = \ab(\partial_{\delta x},\partial_{\delta y},\partial_{\delta z})$. As a result, the total time-averaged acoustic/electromagnetic radiation force $\overline{\bold{F}}$ on the volume $V_p$ in the direction of a small shift $\delta \bold{r}$ is
\begin{equation} \label{eq:Fdeltar}
    \overline{\bold{F}} \cdot \delta \bold{r} = - \frac{\jj}{4 \omega}  (\bold{u}^\dagger \delta \bold{i} + \bold{i}^\dagger \delta \bold{u}).
\end{equation}
We report below how this relation enables unprecedented control over local radiation forces in complex media.

\paragraph{Optimal motion of one object---}
From \cref{eq:immittance_def},
\begin{subequations}
\begin{empheq}[right={\quad.}]{align}
    &  \delta \mathbf{u}=\sqrt{2} \ab[\ab(\mathds{1}+\bold{S}) \delta \bold{s}_+ + \delta \bold{S} \, \bold{s}_+ ] \\
    & \delta \mathbf{i}= \sqrt{2} \ab[\ab(\mathds{1}-\bold{S}) \delta \bold{s}_+ - \delta \bold{S} \, \bold{s}_+ ]
\end{empheq}
\end{subequations}
Since $\delta \bold{s}_+=\bold{0}$ for a fixed chosen input state,  the change in immittance vectors is only due to the parametric change of the $\bold{S}$-matrix: $\bold{u}^\dagger \delta \bold{i} + \bold{i}^\dagger \delta \bold{u} = - 4 \bold{s}_+^\dagger \bold{S}^\dagger \delta \bold{S} \, \bold{s}_+$. This suggests extending the GWS matrix defined in \cref{eq:GWS} to multidimensional parameters $\delta \bold{r}$ as
\begin{equation}
     \bold{Q}_{\delta \bold{r}} = \jj \bold{S}^\dagger \ab(\hat{\bold{e}}_{\bold{r}} \cdot \nabla_{\!\delta \bold{r}})\bold{S} = \frac{\delta x \bold{Q}_{\delta x} + \delta y \bold{Q}_{\delta y} + \delta z \bold{Q}_{\delta z}}{|\delta \bold{r}|},
\end{equation}
where $\hat{\bold{e}}_{\bold{r}}=\delta \bold{r}/|\delta \bold{r}|$. The expectation value over the input states is
\begin{equation}
    \braket* [1] {\bold{Q}_{\delta \bold{r}}}_{\bold{s}_+} \equiv \bold{s}_+^\dagger \bold{Q}_{\delta \bold{r}} \bold{s}_+ = - \frac{\jj}{4  \ab| \delta \bold{r}|} \ab(\bold{u}^\dagger \delta \bold{i} + \bold{i}^\dagger \delta \bold{u}).
\end{equation}
Using \cref{eq:Fdeltar}, we finally obtain
\begin{equation} \label{eq:main_res}
     \frac{1}{\omega} \braket* [1] {\bold{Q}_{\delta \bold{r}}}_{\bold{s}_+} = \overline{\bold{F}} \cdot \hat{\bold{e}}_{\bold{r}}. 
\end{equation}
\cref{eq:main_res} is the central equation of our Letter. It shows that each force component is encoded in how the scattering changes under an infinitesimal shift of the object. Since $\bold{Q}_{\delta \bold r}$ is Hermitian for a unitary $\bold{S}$-matrix (lossless propagation), the projection of the radiation force can be written as a Rayleigh quotient. The Courant–Fischer min–max theorem then ensures that  
\begin{equation}
\frac{\lambda_{1}}{\omega} \ab|\bold{s}_+|^2 \leq \overline{\bold{F}} \cdot \hat{\bold{e}}_{\bold{r}} \leq \frac{\lambda_{N}}{\omega} \ab|\bold{s}_+|^2,
\end{equation}
where the eigenvalues of $\bold{Q}_{\delta \bold{r}}$ are ordered as $\lambda_{1}\leq \lambda_{2}\leq \cdots \leq \lambda_{N}$. Of particular interest are the extremal values of $\overline{\bold{F}} \cdot \hat{\bold{e}}_{\bold{r}}$, which are reached when the incident state coincides with the associated eigenvector, enabling the maximization or minimization of the applied force projection. More generally, if only $L$ out of $N$ ports are accessible, the largest force projection lies between $\lambda_L/\omega$ and $\lambda_N/\omega$. In the following, all the expectation values $\braket* [1]{\cdot}$ are taken with respect to the normalized input state such that $\ab|\bold{s}_+|^2 = 1 \si{\watt}$.

\paragraph{Optimal motion of multiple objects---}

When $N_d$ objects are displaced, the variational integrals split into non-overlapping contributions over the disjoint volumes $V_{p,i}$. The total quantity
\begin{equation} \label{eq:Fdeltar_tot}
    \frac{1}{\omega} \braket* [1] {\bold{Q}_{\delta \bold{r}_\text{tot}}} = \frac{1}{\ab|\delta \bold{r_\text{tot}}|}  \sum_{i=1}^{N_d} \ab|\delta \bold{r}_i| \,\overline{\bold{F}}_i \cdot \hat{\bold{e}}_{\bold{r}_i}
\end{equation}
therefore measures the infinitesimal (virtual) work of the radiation force on the system, normalized by the stacked displacement vector $\delta \bold{r}_\text{tot} = \ab(\delta \bold{r}_1, \cdots, \delta \bold{r}_{N_d})$. $\overline{\bold{F}}_i \cdot \hat{\bold{e}}_{\bold{r}_i}$ is the time-averaged force on the $i$\textsuperscript{th} object in the direction of its push. Without loss of generality, we can focus on the generic case of two objects ($N_d=2$). Because $\bold{Q}_{\delta \bold{r}_1}$ and $\bold{Q}_{\delta \bold{r}_2}$ typically do not commute, they cannot, in general, be diagonalized simultaneously; hence, no single incident state maximizes both projections $\overline{\bold{F}}_1 \cdot \hat{\bold{e}}_{\bold{r}_1}$ and $\overline{\bold{F}}_2 \cdot \hat{\bold{e}}_{\bold{r}_2}$ \footnote{An exception occurs if the two matrices happen to share the same eigenvector associated with their maximal eigenvalues. This non-generic case is understood in \protect \cref {eq:SR-Pareto} by observing that what matters is not ultimately the commutation of the two matrices but the vanishing of the expectation value of the commutator. Further details on the commutation of $\bold{Q}$-matrices, as well as an example of optimization under mirror symmetry, are given in Supplemental Material \cite{supplemental}.}. Yet, the best eigenvector of $\bold{Q}_{\delta \bold{r}_\text{tot}}=(\ab|\delta \bold{r}_1|\bold{Q}_{\delta \bold{r}_1}+\ab|\delta \bold{r}_2|\bold{Q}_{\delta \bold{r}_2})/\ab|\delta \bold{r_\text{tot}}|$ maximizes the sum of the projections. Consequently, the multi-object formulation can be directly interpreted through the lens of multi-objective optimization \cite{miettinenNonlinearMultiobjectiveOptimization1999}. Each $\overline{\bold{F}}_i \cdot \hat{\bold{e}}_{\bold{r}_i}$ defines an objective to be maximized and $\braket* [1] {\bold{Q}_{\delta \bold{r}_\text{tot}}}$ corresponds to a so-called linear scalarization of this problem (i.e., forming a weighted sum to be maximized), with the coefficients $\ab|\delta \bold{r}_i|$ playing the role of optimization weights. An extension of the theory to optimal torque on objects is proposed in the Appendix B. For two objectives -- such as displacing/rotating two objects along/about one direction each -- the feasible set of force components corresponds to the joint numerical range of two Hermitian matrices, which is always convex \cite{hausdorffWertvorratBilinearform1919}. This is also true for three objectives when the number of accessible ports is $N>2$ \cite{fan1987generalized}. In these cases, it is known that linear scalarization generates an entire Pareto front \cite{miettinenNonlinearMultiobjectiveOptimization1999} -- that is, the set of solutions where no objective can be improved without degrading another. We show here that the GWS explores the full set of such compromises. For four or more objectives, convexity can fail, and the GWS recovers only the exposed subset of the Pareto front, leaving non-convex regions inaccessible. 

\begin{figure*}
    \centering
    \includegraphics{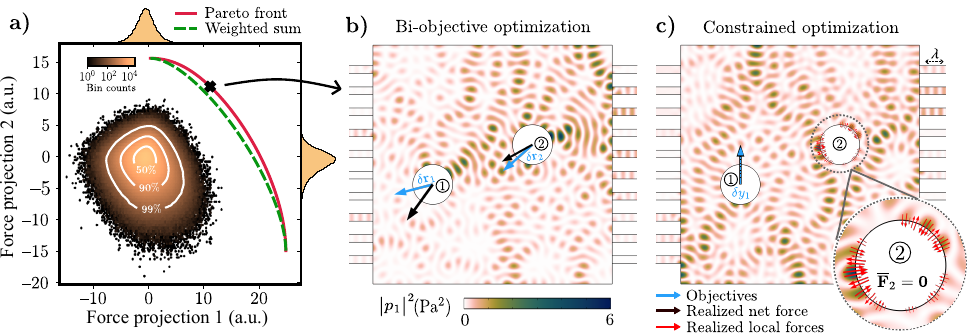}
    \caption{Collective and selective manipulation of objects in scattering media. \textbf{a)} Bi-objective optimization: probability distribution of the force projections on two objects for $10^7$ random input states. The marginal distributions form B-splines. The Pareto front (red solid line) is compared to a heuristic weighted sum of eigenstates (green dashed line).  \textbf{b)} Full wave simulation of the squared acoustic pressure field corresponding to the Pareto-optimal solution for which the force projections are equal, i.e., $\braket* [1] {\bold{Q}_{\delta \bold{r}_1}}=\braket* [1] {\bold{Q}_{\delta \bold{r}_2}}$. The blue arrows are the objectives and the black arrows are the net forces exerted on the objects. \textbf{c)} Constrained optimization: the object \protect\circled{1} is displaced along the vertical direction only (net force in black perfectly overlapping with the objective in blue), while the object \protect\circled{2} experiences local forces (small red arrows with their width and length logarithmically scaling with the local force magnitude) which globally compensate to obtain zero net force. The wavelength $\varlambda$ is shown on the right.}
    \label{fig:subfigs}
\end{figure*}

As an example, a bi-objective Pareto front for the simultaneous translation of two objects in a 20-port disordered 2D acoustic cavity is simulated in \cref{fig:subfigs}a (red solid line).
For comparison, we also show the corresponding force projections for $10^7$ random inputs represented by the distribution of dots, the color of which indicates the number of occurrences. Strikingly, the Pareto front obtained from the GWS procedure greatly outperforms all the random inputs, thus demonstrating that naive Monte Carlo exploration of the input space cannot realistically be implemented to tailor radiation forces in complex media. In the Supplemental Material \cite{supplemental}, we conjecture that the force projections -- whose respective probability distributions are B-splines \cite{gallayNumericalMeasureComplex2012,dunklNumericalShadowsMeasures2011,camposvenutiProbabilityDensityQuantum2013,wangProbabilityDistributionCoherent2025} -- can never compete with the GWS procedure for a sufficiently large number of ports. Moreover, the Pareto front is compared to the result of a heuristic procedure also based on the GWS matrix, where a linear combination of eigenvectors of $\bold{Q}_{\delta \bold{r}_1}$ and $\bold{Q}_{\delta \bold{r}_2}$ is performed (green dashed line). In terms of maximum force projections, the new procedure consistently outperforms this benchmark technique \cite{orazbayevWavemomentumShapingMoving2024}. This demonstrates that our approach can improve existing manipulation schemes. The result of the Pareto optimization is shown in \cref{fig:subfigs}b, where blue arrows correspond to the objectives, and black arrows correspond to the net forces exerted on the objects, computed via the Brillouin stress tensor. We see that the optimized wavefield yields net forces in the intended directions, which confirms that our framework translates directly into the mechanical action predicted by the Brillouin stress tensor -- validating that it is possible to shape the local momentum of waves for complex tweezing scenarios from far-field actuation only. Further examples can be found in the Supplemental Material \cite{supplemental}.

\paragraph{Uncertainty relation---}
In bi-objective optimization with GWS matrices, one seeks to simultaneously maximize the expectation values of the matrices $\bold{Q}_\alpha$ and $\bold{Q}_\beta$, corresponding for instance to distinct forces or torques on one or multiple objects. The central question is how incompatible the objectives can be, and what bound captures the resulting trade-off. To answer this, we define the deficits from utopia values (highest eigenvalues) for a given input state $\bold{s}_+$ as
\begin{equation}
    \epsilon_\alpha = \lambda_N (\bold{Q}_\alpha)  - \braket* [1]{\bold{Q}_\alpha}, \quad \epsilon_\beta = \lambda_N (\bold{Q}_\beta)  - \braket* [1]{\bold{Q}_\beta}.
\end{equation}
A direct application of the Schrödinger-Robertson uncertainty relation \cite{schrodinger1999heisenberg},
combined with the Bhatia-Davis inequality \cite{bhatia2000better},
yields the following bound:
\begin{equation}
\epsilon_\alpha \epsilon_\beta
\geq
\frac{\frac{1}{4}\ab|\langle[\bold{Q}_\alpha,\bold{Q}_\beta]\rangle|^{2} + 
\mathrm{Cov}(\bold{Q}_\alpha, \bold{Q}_\beta)^2}
{(\braket* [1]{\bold{Q}_\alpha}-\lambda_1(\bold{Q}_\alpha)) (\braket* [1]{\bold{Q}_\beta}-\lambda_1(\bold{Q}_\beta))}.
\label{eq:SR-Pareto}
\end{equation}
Here, the symmetrized covariance is
\begin{equation}
    \mathrm{Cov}(\bold{Q}_\alpha, \bold{Q}_\beta)=\frac{1}{2}\ab\langle \ab\{\bold{Q}_\alpha-\braket* [1]{\bold{Q}_\alpha}, \bold{Q}_\beta-\braket* [1]{\bold{Q}_\beta} \}\rangle,
\end{equation}
and $[\bold{A},\bold{B}]$ and $\{\bold{A},\bold{B}\}$ denote the commutator and anti-commutator, respectively. The right-hand side of \cref{eq:SR-Pareto} cannot vanish unless the expectation value of the commutator vanishes and the two GWS matrices are uncorrelated in the chosen input state. In particular, the case $\epsilon_\alpha=\epsilon_\beta=0$ requires the two matrices to share the same eigenstate associated with their respective maximal eigenvalues, a highly non-generic condition. \cref{eq:SR-Pareto} therefore provides a rigorous state-dependent witness of incompatibility: driving one objective closer to its optimum generically restricts how closely the other can approach its own. The Pareto front then identifies the optimal achievable compromises within this constrained landscape, while the corresponding GWS solutions generate strongly directed forces that are exceptionally unlikely to arise from random inputs.

\paragraph{Constrained optimization---}
So far, our theory allows us to maximize multiple objectives simultaneously. For example, in \cref{fig:subfigs}b, the wavefront is the best one to move both objects along the blue arrows. However, the optimal forces are not exactly in the desired directions, because the orthogonal force components are left unconstrained. A natural question we ask now is: can we generalize the framework to handle additional constraints, such as no motion of certain objects in particular directions? Formally, consider the problem of maximizing the expectation value of a given $\bold{Q}_\alpha$ while enforcing $N_c$ constraints on the expectation value of other objectives $\bold{Q}_{\beta_i}$. This is known as a complex-valued homogeneous quadratically constrained quadratic program, and is generally an NP-hard problem for $N_c>2$ \cite{huangRankConstrainedSeparableSemidefinite2010,luoSemidefiniteRelaxationQuadratic2010}. However, established algorithms can easily find local optima, allowing us to include constraints within our theory. As an example, we show in \cref{fig:subfigs}c the manipulation of object \circled{1} in the $y$-direction only (constraining the $x$-motion to zero), while object \circled{2} is kept completely still, i.e., $\braket* [1]{\bold{Q}_{\delta y_1}}$ is maximized, while $\braket* [1]{\bold{Q}_{\delta x_1}}$, $\braket* [1]{\bold{Q}_{\delta x_2}}$ and $\braket* [1]{\bold{Q}_{\delta y_2}}$ are forced to be close to zero. The Gurobi Optimization solver is used for this task \cite{gurobi}. The resulting wave field indeed manages to compensate local forces acting on an object, as indicated by the red arrows in the inset of \cref{fig:subfigs}c, and cancels the prescribed force projections (more details in Supplemental Material \cite{supplemental}). This represents another multi-objective tweezing task of high practical relevance. 

\paragraph{Discussion---}
We have shown that local radiation forces on targets embedded in complex scattering media can be optimized through the GWS matrix: its extremal eigenstates maximize force or torque projections on a single object, while in the multi-object case they realize Pareto-optimal trade-offs among competing objectives. Together with a realistic demonstration of constrained optimization of objects' motion along prescribed directions, our Letter establishes a unified framework across acoustics and electromagnetism, and provides a principled route to coordinated manipulation in strongly scattering environments. In addition, the uncertainty relation of \cref{eq:SR-Pareto} quantifies the fundamental trade-offs that arise when simultaneously optimizing force or torque components and shows that Pareto optimization in wave momentum shaping mirrors the mathematical structure associated with incompatible quantum observables \cite{schrodinger1999heisenberg}.

\begin{figure}
    \centering
    \includegraphics{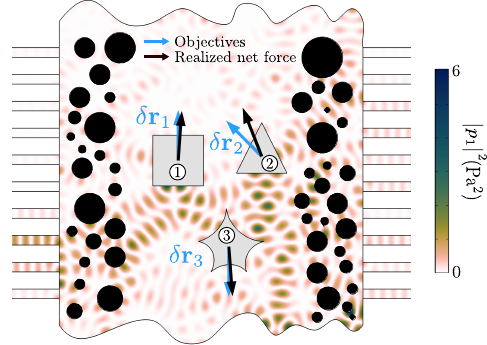}
    \caption{Example of tri-objective optimization through a turbid medium (fixed black circular scatterers). A single incident wavefront is optimized to simultaneously impose independent radiation force objectives on three targets of arbitrary shape (in gray: square, triangle, and star). The background colormap shows the resulting squared pressure field; arrows indicate the prescribed force directions (blue) and the corresponding realized net forces on each object (black).}
    \label{fig:fig3}
\end{figure}

Beyond fundamental interest, our results may contribute to the development of in vivo precision therapy and drug delivery, where navigation through deep tissues requires controlling radiation forces in turbid media \cite{favre2019optical,del_campo_fonseca_ultrasound_2024,ghanemNoninvasiveAcousticManipulation2020,loTornadoinspiredAcousticVortex2021a,delcampofonsecaUltrasoundTrappingNavigation2023a,yangInvivoProgrammableAcoustic2023,medanyModelbasedReinforcementLearning2025,burstowEvaluatingAccuracyAcoustic2025}. From a broader perspective, our framework applies across wave platforms and frequency ranges, from acoustofluidic manipulation to electromagnetic actuation. As a final illustration of the generality of the method, \cref{fig:fig3} shows how a single incident wavefront can simultaneously enforce multiple independent radiation force objectives on multiple targets of arbitrary geometry, despite the presence of a turbid scattering medium.
We also note interesting connections between our findings and a recent article \cite{liJointControlCoherent2025}, which investigates the simultaneous control of transmission, reflection and absorption. Our Letter strongly suggests that the notion of Pareto-optimality for multi-objective wavefront shaping introduced here can serve as a common framework for such joint-control scenarios beyond forces and torques.

An important next step is the experimental validation of our multi-objective predictions, including direct tests of Pareto fronts and constrained actuation in realistic multi-scatterer environments. A second direction is the theoretical extension to lossy media, for which the unitary foundation established here provides the natural and rigorous starting point. Promisingly, experiments already indicate that protocols based on GWS matrices remain highly effective in practice even in the presence of absorption \cite{orazbayevWavemomentumShapingMoving2024}. Together, these developments open a clear path toward the ultimate goal: collective and selective manipulation of multiple objects in lossy, strongly scattering media.

\paragraph{Acknowledgments---} This work was supported by the Swiss National Science Foundation (SNSF) under the grant No. 10001567.

\paragraph{Data Availability---} The data that support the findings of this article are openly available \cite{zenodo}.

\paragraph{Appendix A: Derivation details for \cref{eq:main_varia}---}
We show here how \cref{eq:main_varia} is derived. It extends an important variational integral relation \cite{hupfl2023optimal,ren2022rigorous} connecting parametric changes occurring in the scattering region to changes of the immittance voltages and currents in the asymptotic regions. Defining the complex Poynting vector as
\begin{subequations}
\begin{empheq}[left={\bold{\Pi} = \empheqlbrace}]{align}
    & \frac{1}{2} p_1 \bold{v}_1^* \quad \text{(in acoustics)},\\
    & \frac{1}{2}\bold{E} \times \bold{H}^* \quad \text{(in electromagnetism)},
\end{empheq}
\end{subequations}
it can be shown that in both acoustics and electromagnetism \cite{barybinModalExpansionsOrthogonal1998} 
\begin{equation} \label{eq:immittance_integral}
    - 2 \int_A \ab(\bold{\Pi} + \bold{\Pi}^*) \cdot \diff{\mathbf{A}}  = \mathbf{u^\dagger} \mathbf{i} + \mathbf{i}^\dagger\mathbf{u}
\end{equation}
holds, where $\bold{A}$ is the boundary of the scattering volume $V$, oriented away from the scattering region. Equation \eqref{eq:immittance_integral} holds whenever no reactive wave enters the volume $V$, i.e., the scattering region is only excited via propagating waves from the outside. Experimentally, this often requires placing sources in the far-field, so that all evanescent waves are sufficiently decayed. For a given physical state $a$, \cref{eq:immittance_integral} reads in acoustics
\begin{equation}
    P(a) \equiv - \int_A \ab(p_{1a} \bold{v}_{1a}^* + p_{1a}^* \bold{v}_{1a}) \cdot \diff{\mathbf{A}}  = \mathbf{u}_a^\dagger \mathbf{i}_a + \mathbf{i}_a^\dagger\mathbf{u}_a.
\end{equation}
By extension, it is possible to compose any other physical state $a\pm b$ or $a \pm \jj b$. Using the linearity of $p_1$, $\bold{v}_1$, $\bold{u}$ and $\bold{i}$, one can verify the so-called polarization identity \cite{blanchard2015inner}: 
\begin{equation}\label{eq:polarization}
    \begin{split}
        &\frac{P(a+b) + \jj P(a+\jj b) - \ab[P(a-b) + \jj P(a-\jj b)]}{4} \\ & \quad =  - \int_A \ab(p_{1a} \bold{v}_{1b}^* + p_{1b}^* \bold{v}_{1a}) \cdot \diff{\mathbf{A}} = \mathbf{u}_b^\dagger \mathbf{i}_a + \mathbf{i}_b^\dagger\mathbf{u}_a.
    \end{split}
\end{equation}
Taking the complex conjugate of \cref{eq:polarization} for the states $a\rightarrow (p_1, \bold{v}_1,\bold{u},\bold{i})$ and $b\rightarrow (\delta p_1, \delta\bold{v}_1,\delta\bold{u},\delta\bold{i})$, we obtain
\begin{equation}\label{eq:acous_pola}
    -\int_A  \ab(p_1^* \mathbf{\delta\bold{v}_1}+\delta p_1 \bold{v}_1^*) \cdot \diff{\bold{A}} = \bold{u}^\dagger \delta \bold{i} + \bold{i}^\dagger \delta \bold{u}.
\end{equation}
Now, we use $\delta$ to denote small variations of fields or parameters due to perturbations such that \mbox{$\delta f = \nabla_{\! \delta \bold{\alpha}} f \cdot \delta \bold{\alpha}$}. In acoustics, \begin{equation}\label{eq:acoustics_hamronic_approx_equations}
\nabla \cdot \bold{v}_1 = -\jj \omega \kappa_0 p_1  \text{ and } \nabla p_1 = - \jj \omega \rho_0 \bold{v}_1.
\end{equation}
At first variational order,
\begin{subequations}
\begin{empheq}{align}
    & \nabla \cdot \ab(\delta \bold{v}_1) = -\jj \omega \delta \bold{\alpha} \cdot \nabla_{\scriptscriptstyle{\!\! \delta \bold{\alpha}}} \ab(\kappa_0 p_1), \label{eq:var_ac1}  \\
    & \nabla (\delta p_1) = - \jj \omega \delta \bold{\alpha} \cdot \nabla_{\scriptscriptstyle{\!\! \delta \bold{\alpha}}} \ab(\rho_0 \bold{v}_1). \label{eq:var_ac2}
\end{empheq}
\end{subequations}
To relate this to the immittance vectors, \cref{eq:var_ac1,eq:var_ac2} may be combined as follows:
\begin{equation} \label{eq:ac_nabla_alpha}
    - \nabla \cdot (p_1^* \mathbf{\delta\bold{v}_1}+\delta p_1 \bold{v}_1^*)=\jj \omega \delta \bold{\alpha} \cdot \ab( \ab|p_1|^2 \nabla_{\scriptscriptstyle{\!\! \delta \bold{\alpha}}} \kappa_0 +\ab|\bold{v}_1|^2 \nabla_{\scriptscriptstyle{\!\! \delta \bold{\alpha}}} \rho_0 ).
\end{equation}
It is then possible to compute the volume integral of \cref{eq:ac_nabla_alpha} and use \cref{eq:acous_pola} to obtain
\begin{equation} \label{eq:acous-varia}
    \bold{u}^\dagger \delta \bold{i} + \bold{i}^\dagger \delta \bold{u} = \jj \omega \delta \bold{\alpha} \cdot \int_V \ab(\ab|p_1|^2 \nabla_{\scriptscriptstyle{\!\! \delta \bold{\alpha}}} \kappa_0 +\ab|\bold{v}_1|^2 \nabla_{\scriptscriptstyle{\!\! \delta \bold{\alpha}}} \rho_0 ) \diff V.
\end{equation}
In electromagnetism, Maxwell-Faraday and -Ampère laws in frequency domain give
\begin{subequations}
\begin{empheq}{align}
    & \nabla\times\delta\bold{E}=-\jj\omega  \delta \bold{\alpha} \cdot \nabla_{\scriptscriptstyle{\!\! \delta \bold{\alpha}}} \ab(\mu \bold{H}) \\
    & \nabla\times\delta\bold{H}=\jj\omega \delta \bold{\alpha} \cdot \nabla_{\scriptscriptstyle{\!\! \delta \bold{\alpha}}} \ab(\varepsilon \bold{E})
\end{empheq}
\end{subequations}
and
\begin{equation} \label{eq:em_nabla_alpha}
\begin{split}
    - \nabla \cdot &\ab(\bold{E}^* \times \delta \bold{H} + \delta \bold{E} \times \bold{H}^*) \\ &= \jj \omega \delta \bold{\alpha} \cdot \ab(|\bold{E}|^2 \nabla_{\scriptscriptstyle{\!\! \delta \bold{\alpha}}} \varepsilon + |\bold{H}|^2 \nabla_{\scriptscriptstyle{\!\! \delta \bold{\alpha}}} \mu),
\end{split}
\end{equation}
from which we obtain
\begin{equation} \label{eq:em-varia}
    \bold{u}^\dagger \delta \bold{i} + \bold{i}^\dagger \delta \bold{u} = \jj \omega \delta \bold{\alpha} \cdot \int_V \ab(|\bold{E}|^2 \nabla_{\scriptscriptstyle{\!\! \delta \bold{\alpha}}} \varepsilon + |\bold{H}|^2 \nabla_{\scriptscriptstyle{\!\! \delta \bold{\alpha}}} \mu) \diff V.
\end{equation}
This is completely analogous to the acoustic case and yields the same end result. 

\paragraph{Appendix B: Optimal torque---}
Rotations enter the variational identities \cref{eq:acous-varia,eq:em-varia} through the generator of infinitesimal rigid rotations: a small angle $\delta \bold{\theta}$ about a pivot point $\bold{r}_0$ moves each material point by $\delta \bold{r} = \delta \bold{\theta} \times (\bold{r}- \bold{r}_0)$. This gives the parametric derivative $\nabla_{ \!\delta\bold{\theta}}= - (\bold{r}- \bold{r}_0) \times \nabla$. Defining the time-averaged torque $\overline{\bold{\tau}}(\bold{r}_0)$ as
\begin{equation}
    \overline{\bold{\tau}}(\bold{r}_0) = \int_V (\bold{r}- \bold{r}_0) \times \overline{\bold{f}} \diff{V},
\end{equation}
one obtains
\begin{equation}
    \overline{\bold{\tau}}(\bold{r}_0) \cdot \delta \bold{\theta} = - \frac{\jj}{4 \omega}  \ab(\bold{u}^\dagger \delta \bold{i} + \bold{i}^\dagger \delta \bold{u}).
\end{equation}
Then, in analogy with the linear force case, the GWS reads
\begin{equation}\label{eq:torque}
    \frac{1}{\omega} \braket* [1] {\bold{Q}_{\delta\bold{\theta}}} = \overline{\bold{\tau}}(\bold{r}_0) \cdot \hat{\bold{n}},
\end{equation}
and
\begin{equation}
    \frac{1}{\omega} \braket* [1] {\bold{Q}_{\delta\bold{\theta}_\text{tot}}^{(\bold{r}_{0,i})}} = \frac{1}{\ab|\delta  \bold{\theta_\text{tot}}|}  \sum_{i=1}^{N_d} \ab|\delta \bold{\theta}_i| \overline{\bold{\tau}}_i(\bold{r}_{0,i}) \cdot \hat{\bold{n}}_i,
\end{equation}
where $\hat{\bold{n}}=\delta \bold{\theta}/|\delta \bold{\theta}|$. Interestingly, for axisymmetric objects around $\bold{\theta}$ (such as the discs in \cref{fig:subfigs}), $\bold{Q}_{\delta\bold{\theta}}=\bold{0}$, and no input state can transfer angular momentum to the objects. This counter-intuitive result, holding for any lossless scattering, was already discovered in both acoustics and electromagnetism via other methods, see for example Refs.~\cite{toftulRadiationForcesTorques2025,marstonRadiationTorqueSphere1984,zhangAngularMomentumFlux2011,silvaRadiationTorqueProduced2012,smaginAcousticLateralRecoil2024,chaumetElectromagneticForceTorque2009,leeOpticalTorqueEnhanced2014,nieto-vesperinasOpticalTorqueSmall2015}. 


%

\begingroup
\parskip=\baselineskip
\parindent=0pt
\onecolumngrid{
\section{Supplemental Material for ``Multi-Objective Tweezers in Scattering Media''}

\subsection{Scattering amplitudes and acoustic/electromagnetic fields in multimode waveguides.}

For simplicity, we choose in the main text the example of multiport cavities with monomode waveguides as the input/output leads. In general, the $\bold{S}$-matrix formalism also holds for free space scattering, and for any arrangement of waveguides, including multimode ones. Below, we show the exact link between the scattering amplitudes and the acoustic/electromagnetic fields in these waveguides.  

\textbf{Acoustic case.} 

The acoustic pressure field in a waveguide (here directed along $z$) of cross-section $A$ admits the modal expansion
\begin{equation}
p_1(x,y,z)=\sum_m a_m \,\chi_m(x,y)\,\ee^{\,\jj(\omega t-k_m z)},
\end{equation}
where $\chi_m(x,y)$ is the (dimensionless) transverse profile of the $m$\textsuperscript{th} mode. We adopt the normalization
\begin{equation}\label{eq:Acnorm}
\int_A \diff{S}\,\ab|\chi_m(x,y)|^2 = A,
\end{equation}
so that the complex amplitudes $a_m$ carry the units of pressure. Introducing power-normalized scattering amplitudes $s_m^\pm$ such that $\ab|s_m^\pm|^2$ equals the incident ($+$) or outgoing ($-$) power in port/mode $m$, the acoustic Poynting vector yields the equivalent representation
\begin{equation}\label{eq:mastereqacoustics}
p_1(x,y,z)=\sum_m \sqrt{\frac{2\omega \rho_0}{\ab|k_m|\, A}}\; s^\pm_m \,\chi_m(x,y)\,\ee^{\,\jj(\omega t-k_m z)}.
\end{equation}
Here the choice of superscript is tied to the sign of the longitudinal wavenumber: use $s_m^{+}$ when $k_m>0$ and $s_m^{-}$ when $k_m<0$. This convention assumes that, in each port, the local coordinate $z$ is oriented toward the scattering region (e.g., for a port on the left of the scattering region, $z$ increases to the right). Accordingly, for each connected waveguide, $z$ should be understood as the local axial coordinate pointing toward the scattering region.

\textbf{Electromagnetic case accounting for polarization.}

In an electromagnetic waveguide of cross-section $A$, the fields admit a polarized modal expansion into traveling TE/TM modes (See standard textbooks, such as Ref.~\cite{chew2021lectures}),
\begin{align}
\mathbf E^{\mathrm{TE}}(x,y,z) &= \jj\,\ee^{\,\jj\omega t}\sum_m
\sqrt{\frac{2\omega\mu}{\ab|k_m|}}\;
\big(\hat{\mathbf z}\times\nabla_t h^{\mathrm{TE}}_m(x,y)\big)\,
\Big(s^{+}_{\mathrm{TE},m}\ee^{-\jj k_m z}+s^{-}_{\mathrm{TE},m}\ee^{+\jj k_m z}\Big), \label{eq:EM_TE_E}\\
\mathbf H^{\mathrm{TE}}(x,y,z) &= \ee^{\,\jj\omega t}\sum_m\Bigg[
\hat{\mathbf z}\,\frac{k_{t,m}^2\,h^{\mathrm{TE}}_m(x,y)}{\sqrt{\omega\mu\,\ab|k_m|/2}}\,
\Big(s^{+}_{\mathrm{TE},m}\ee^{-\jj k_m z}+s^{-}_{\mathrm{TE},m}\ee^{+\jj k_m z}\Big)
\\ & \qquad\qquad\qquad -\jj\sqrt{\frac{2\ab|k_m|}{\omega\mu}}\;\nabla_t h^{\mathrm{TE}}_m(x,y)\,
\Big(s^{+}_{\mathrm{TE},m}\ee^{-\jj k_m z}-s^{-}_{\mathrm{TE},m}\ee^{+\jj k_m z}\Big)\Bigg], \notag \\[1mm]
\mathbf H^{\mathrm{TM}}(x,y,z) &= -\jj\,\ee^{\,\jj\omega t}\sum_m
\sqrt{\frac{2\omega\varepsilon}{\ab|k_m|}}\;
\big(\hat{\mathbf z}\times\nabla_t e^{\mathrm{TM}}_m(x,y)\big)\,
\Big(s^{+}_{\mathrm{TM},m}\ee^{-\jj k_m z}+s^{-}_{\mathrm{TM},m}\ee^{+\jj k_m z}\Big), \label{eq:EM_TM_H}\\
\mathbf E^{\mathrm{TM}}(x,y,z) &= \ee^{\,\jj\omega t}\sum_m\Bigg[
\hat{\mathbf z}\,\frac{k_{t,m}^2\,e^{\mathrm{TM}}_m(x,y)}{\sqrt{\omega\varepsilon\,\ab|k_m|/2}}\,
\Big(s^{+}_{\mathrm{TM},m}\ee^{-\jj k_m z}+s^{-}_{\mathrm{TM},m}\ee^{+\jj k_m z}\Big)
\\ & \qquad\qquad\qquad -\jj\sqrt{\frac{2\ab|k_m|}{\omega\varepsilon}}\;\nabla_t e^{\mathrm{TM}}_m(x,y)\,
\Big(s^{+}_{\mathrm{TM},m}\ee^{-\jj k_m z}-s^{-}_{\mathrm{TM},m}\ee^{+\jj k_m z}\Big)\Bigg]. \notag
\end{align}
Here $k_{t,m}^2=\omega^2\mu\varepsilon-k_m^2$, $\nabla_t$ is the transverse gradient, and the longitudinal mode profiles are taken dimensionless and normalized as
\begin{equation}
\int_A \diff{S}\,\ab|\nabla_t h^{\mathrm{TE}}_m|^2=\int_A \diff{S}\,\ab|\nabla_t e^{\mathrm{TM}}_m|^2=1.
\end{equation}
The complex scattering amplitudes $s^{\pm}_{\mathrm{TE},m}$ and $s^{\pm}_{\mathrm{TM},m}$ are power-normalized so that $\ab|s^{\pm}_m|^2$ equals the incident ($+$) or outgoing ($-$) time-averaged power in the corresponding polarized mode. The superscript is tied to the sign convention for $k_m$ with the local port coordinate $z$ oriented toward the scattering region.

It can be readily seen from the above relations between the scattering amplitudes and the actual TE/TM mode amplitudes that the framework naturally accounts for polarization by treating TE and TM as additional channel labels, i.e., by enlarging the modal basis and letting the scattering vectors and $\bold{S}$-matrix act on this augmented channels space (with any TE-TM conversion appearing as off-diagonal blocks).

\subsection{Commutation of $\bold{Q}$-matrices and optimization under mirror (reflection) symmetry}

When $[\mathbf{Q}_\alpha, \mathbf{Q}_\beta]=0$, the two objectives are algebraically compatible: the operators can be simultaneously diagonalized (i.e.\ they share the same eigenstates), and the uncertainty-type limitation associated with non-commutation becomes trivial when choosing a common eigenstate. As stated in the main text, simultaneous maximization of both objectives is then possible \emph{if} the two operators share the same \emph{maximizing} eigenstate. A trade-off can persist if the respective maxima occur on different common eigenstates; in that situation the Pareto front is simply obtained by varying the contributions of these shared eigenstates rather than by an uncertainty-type bound. Geometrically, for two objectives the shared eigenstates define points in objective space, and the Pareto front is the outer chain of line segments connecting the relevant ones.

Physically, a vanishing commutator of the two GWS matrices can arise when the corresponding objectives do not compete for the same scattering degrees of freedom, either because they act on effectively decoupled subspaces (e.g., negligible cross-coupling between two objects), or because symmetry separates the problem into invariant sectors (e.g., even/odd under mirror reflection in a mirror-symmetric system). These cases are typically non-generic in non-symmetric and lossless complex scattering media, since such systems usually induce strong mode mixing.  

Mathematically, \emph{if} a lossless scattering matrix $\mathbf{S}$ can be written in one and the same \emph{parameter-independent} modal basis as a fixed block-diagonal matrix for all parameter values (i.e., the modes split into invariant sectors that never couple), then every GWS matrix is block-diagonal in exactly the same basis, so parameter changes cannot transfer amplitude between different blocks. Concretely, in that basis one has
\begin{equation}
\mathbf{Q}_\alpha=
\begin{pmatrix}
\mathbf{Q}_{\alpha,1}&0&\cdots&0\\
0&\mathbf{Q}_{\alpha,2}&\cdots&0\\
\vdots&\vdots&\ddots&\vdots\\
0&0&\cdots&\mathbf{Q}_{\alpha,N}
\end{pmatrix},
\qquad
\mathbf{Q}_\beta=
\begin{pmatrix}
\mathbf{Q}_{\beta,1}&0&\cdots&0\\
0&\mathbf{Q}_{\beta,2}&\cdots&0\\
\vdots&\vdots&\ddots&\vdots\\
0&0&\cdots&\mathbf{Q}_{\beta,N}
\end{pmatrix},
\end{equation}
so the commutator is itself block-diagonal,
\begin{equation}
[\mathbf{Q}_\alpha,\mathbf{Q}_\beta]=
\begin{pmatrix}
[\mathbf{Q}_{\alpha,1},\mathbf{Q}_{\beta,1}]&0&\cdots&0\\
0&[\mathbf{Q}_{\alpha,2},\mathbf{Q}_{\beta,2}]&\cdots&0\\
\vdots&\vdots&\ddots&\vdots\\
0&0&\cdots&[\mathbf{Q}_{\alpha,N},\mathbf{Q}_{\beta,N}]
\end{pmatrix},
\end{equation}
and therefore $\mathbf{Q}_\alpha$ and $\mathbf{Q}_\beta$ commute overall exactly when they commute inside every block (equivalently, a single nonzero block commutator makes the full commutator nonzero). In our cases, these blocks are groups of scattering modes that share a conserved label -- for example even/odd sectors in a mirror-symmetric structure, fixed azimuthal index sectors in an axisymmetric scatterer, TE/TM polarization sectors when polarization mixing is forbidden, or decoupled sub-port subspaces in a network with negligible cross-coupling. Symmetry typically guarantees this inter-block decoupling, but within any block of dimension $>1$ the system can still mix the modes in that sector, so commutation within that block is not automatic and depends on whether the two matrices correspond to compatible (simultaneously diagonalizable) mode transformations inside that sector. In particular, if the blocks are one-dimensional (so $\mathbf{S}$ is diagonal in that fixed basis), then all $\mathbf{Q}_\alpha$ are diagonal in that same basis and therefore commute pairwise.

The mirror-symmetric two-port example below is a simple illustration of this principle: mirror (reflection) symmetry enforces a parameter-independent even/odd block decomposition of $\mathbf{S}$, so the associated $\mathbf{Q}_\alpha$ and $\mathbf{Q}_\beta$ inherit the same block structure and the objectives commute within that symmetry-protected basis.

\textbf{Example: 2-port unitary toy model with mirror symmetry.} 

Consider the following scattering matrix, describing a simple lossless and spatially symmetric 2-port system with left (L) and right (R) ports:
\begin{equation}
\mathbf{S}^\text{RL} = \begin{pmatrix}
        r & t \\ t & r
    \end{pmatrix},
\end{equation}
such that
\begin{equation}
    \begin{pmatrix}
        s_-^\text{L} \\ s_-^\text{R}
    \end{pmatrix} = \mathbf{S}^\text{RL} \begin{pmatrix}
        s_+^\text{L} \\ s_+^\text{R}
    \end{pmatrix}. 
\end{equation}
Defining the even (E) and odd (O) basis under left-right reflection as
\begin{equation}
    s_\pm^\text{E} = \frac{s_\pm^\text{L}+s_\pm^\text{R}}{\sqrt{2}} \quad \text{and} \quad s_\pm^\text{O} = \frac{s_\pm^\text{L}-s_\pm^\text{R}}{\sqrt{2}},
\end{equation}
one can write
\begin{equation}
    \begin{pmatrix}
        s_-^\text{E} \\ s_-^\text{O}
    \end{pmatrix} = \mathbf{S}^\text{EO} \begin{pmatrix}
        s_+^\text{E} \\ s_+^\text{O}
    \end{pmatrix}
\end{equation}
with
\begin{equation}
\mathbf{S}^\text{EO} = \begin{pmatrix}
        r + t & 0 \\ 0 & r-t
    \end{pmatrix} = \begin{pmatrix}
    \mathrm{e}^{\,\mathrm{j} \phi_\text{E}} & 0 \\ 0 & \mathrm{e}^{\,\mathrm{j} \phi_\text{O}}
    \end{pmatrix}.
\end{equation}
We can see that the EO basis, which is the same for all parameters, symmetrizes the scattering matrix. Then, the GWS matrices with respect to $\alpha$ and $\beta$ in this fixed basis are both diagonal: \begin{equation}
    \mathbf{Q}_\alpha^\text{EO} = - \begin{pmatrix}
    \partial_\alpha \phi_\text{E}  & 0 \\ 0 & \partial_\alpha\phi_\text{O}
    \end{pmatrix} \quad \text{and} \quad \mathbf{Q}_\beta^\text{EO} = -\begin{pmatrix}
    \partial_\beta \phi_\text{E}  & 0 \\ 0 & \partial_\beta\phi_\text{O}
    \end{pmatrix},
\end{equation}
such that $[\mathbf{Q}_\alpha^\text{EO}, \mathbf{Q}_\beta^\text{EO}]=0$. Importantly, note that the commutation relations are invariant under any parameter-independent change of basis, so one can at convenience compute the matrices back in the left / right basis. Now, if for example $\alpha$ and $\beta$ are two force projections on two different objects, one can compute the force projections produced by the even eigenstate and by the odd eigenstate. If both forces are maximized by the same eigenstate, then one can simultaneously maximize them. However, if the two forces are maximized by different eigenstates, then extreme Pareto points are the two eigenvalues, and all achievable solutions -- forming a line segment between the two eigenvalues -- are Pareto-optimal. This is illustrated in Fig.~\ref{fig:parity}.
\begin{figure}
    \centering
\includegraphics{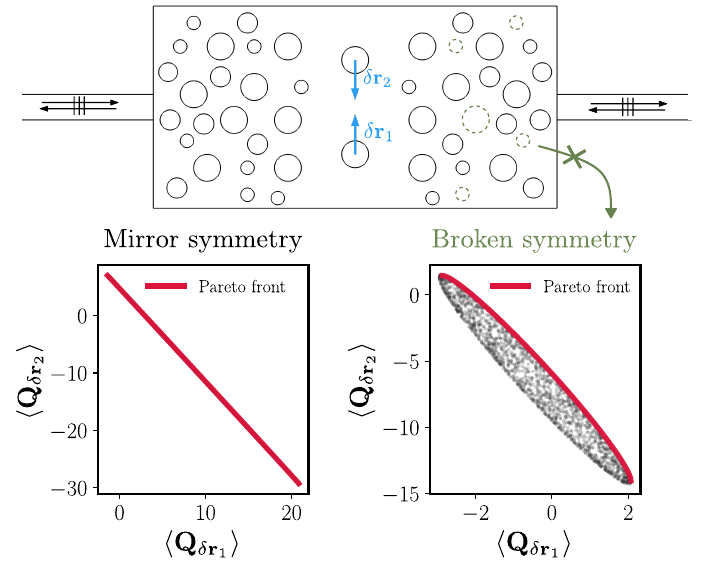}
    \caption{Impact of mirror (reflection) symmetry on bi-objective wave-momentum shaping. Top: Two versions of a 2D disordered scattering region. The system is spatially symmetric when the scatterers are identical on the left and right; the symmetry is broken when the green dashed scatterers are removed. Two objects at the center undergo small translations $\delta\mathbf{r}_1$ and $\delta\mathbf{r}_2$ defining two force-projection objectives. Bottom: The corresponding achievable objective pairs and their Pareto fronts (red): the left plot corresponds to the symmetric case above, and the right plot corresponds to the symmetry-broken one. Results from $2\times 10^3$ random input states are shown in both cases. In the commuting (symmetric) case, all outcomes collapse onto a line segment (the gray dots are completely hidden behind the Pareto front); when the objectives do not commute (symmetry broken), the outcomes fill a two-dimensional region.}
    \label{fig:parity}
\end{figure}

In general, for several $N\times N$ commuting objective matrices, all trade-offs are purely classical: every achievable objective vector is a convex combination of the objective vectors associated with the common eigenstates, and the Pareto-optimal set is the ``upper'' boundary of that convex hull.

\subsection{Statistics of expectation values}

The probability density of expectation values of any $N\times N$ Hermitian matrix $\bold{Q}_\alpha$ over random input states sampled uniformly from the unit sphere in the $N$-dimensional input space, forms a unimodal non-negative B-spline (also called an M-spline) of degree $N-2$, with knots corresponding to the eigenvalues of the matrix:
\begin{equation}
    f_{\braket* [1]{\bold{Q}_\alpha}} (\xi) = (N-1) \sum_{i=1}^N \frac{\max(0,\lambda_i - \xi)^{N-2}}{\prod \limits_{j\neq i} (\lambda_i - \lambda_j)}.
\end{equation}
Examples of this distribution for some random Hermitian matrices are given in \cref{fig:pdf}. The case $N=2$ -- where the probability distribution is uniform -- is equivalent to the one of \cref{fig:parity}.
\begin{figure*}[h]
    \centering
    \includegraphics[width=0.7\linewidth]{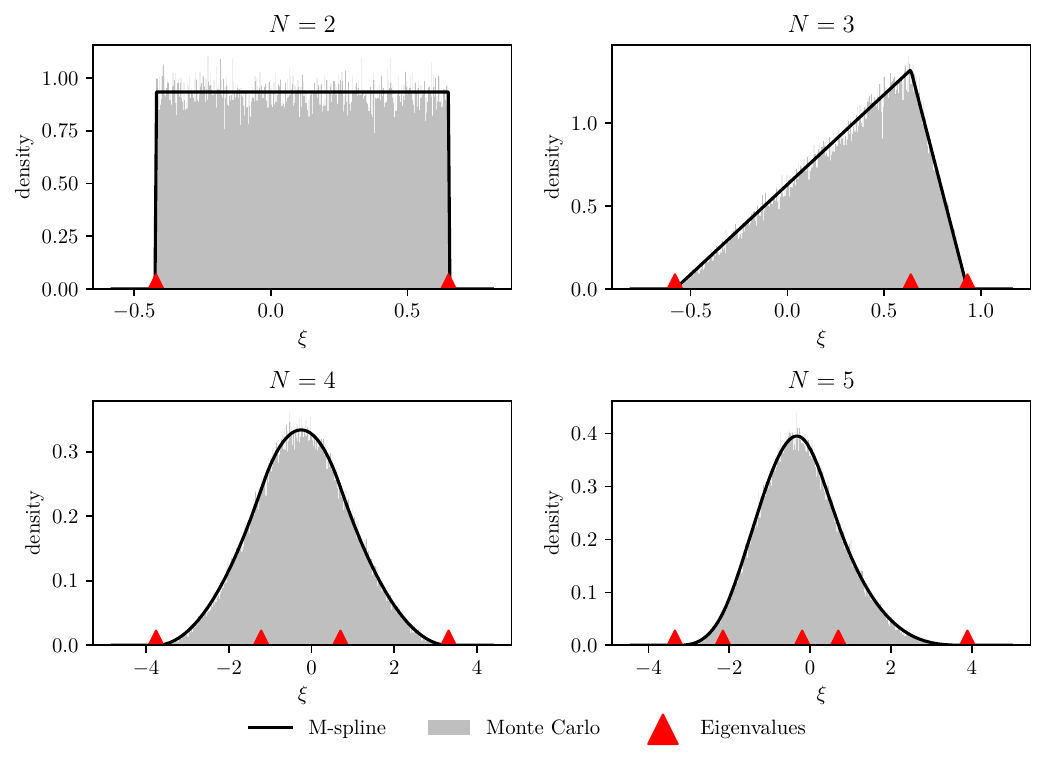}
    \caption{Probability density function of expectation values of Hermitian matrices of size $N$. Comparison between $10^5$ random input states (in grey) and the theoretical M-spline (black curves).}
    \label{fig:pdf}
\end{figure*}

The expected value of $\braket* [1]{\bold{Q}_\alpha}$ is
\begin{equation}
    \mathbb{E}(\braket* [1]{\bold{Q}_\alpha}) = \frac{\operatorname{tr}(\bold{Q}_\alpha)}{N},
\end{equation}
and the variance is
\begin{equation}
    \operatorname{Var}(\braket* [1]{\bold{Q}_\alpha}) = \frac{\dfrac{\operatorname{tr}(\bold{Q}_\alpha^2)}{N}-\ab(\dfrac{\operatorname{tr}(\bold{Q}_\alpha)}{N})^2}{N+1} \leq \frac{\max(|\lambda_1|,|\lambda_N|)^2}{N+1},
\end{equation}
which goes to zero for large $N$, since $|\lambda_1|$ and $|\lambda_N|$ are physically bounded. The probability that the expectation value on a random input state approaches the largest eigenvalue of $\bold{Q}_\alpha$ up to a small fixed parameter $\epsilon$ is
\begin{equation}
    \mathbb{P}(\braket* [1]{\bold{Q}_\alpha} > \lambda_N - \epsilon) = \frac{\epsilon^{N-1}}{\prod\limits_{i=1}^{N-1} (\lambda_N - \lambda_i)} \leq \ab(\frac{\epsilon}{\lambda_N - \lambda_{N-1}})^{N-1}, \quad  \epsilon < \lambda_N - \lambda_{N-1}.
\end{equation}
The upper bound goes to zero for increasing $N$, if
\begin{equation}
    \lim_{N\to + \infty} N \ab(1-\frac{\epsilon}{\lambda_N - \lambda_{N-1}}) = + \infty,
\end{equation}
which is likely to hold, except if the spectrum of $\bold{Q}_\alpha$ shrinks to the top (i.e., $\lambda_{N-1} \to \lambda_N$) sufficiently fast. We conjecture that this is unexpected without any symmetry in the system. It is worth noting that the exact statistics of the distributions of eigenvalues of $\bold{Q}_\alpha$ have been studied by Brouwer, Frahm and Beenakker \cite{brouwer_quantum_1997}, which may provide a good basis for a more in-depth treatment of this likelihood. Analogue discussions were recently performed in the case of the transmissivity distribution by Wang and Guo \cite{wangProbabilityDistributionCoherent2025}.
}

\subsection{Further examples of collective manipulation of objects}

In Fig.~\ref{fig:supp-biobj}, we show further examples of bi-objective optimization; in Fig.~\ref{fig:supp-triobj}, we show further examples of tri-objective optimization for both circular and non circular objects. Blue arrows denote the prescribed objective directions, while black arrows denote the realized Pareto-optimal net forces on the objects. Beyond the angular mismatch, the different arrow lengths also indicate that the achieved force magnitudes are generally unequal and depend on the particular compromise selected on the Pareto front. In the tri-objective case, both the larger directional mismatch and the larger disparity in force magnitudes reflect the fact that the number of available cavity modes was kept fixed while the number of imposed objectives was increased. Increasing the number of ports and/or the mode-mixing capability of the cavity would provide more controllable degrees of freedom and therefore improve independent control over both the directions and amplitudes of the forces.

\begin{figure}[ht]
    \centering
    \includegraphics[width=\linewidth]{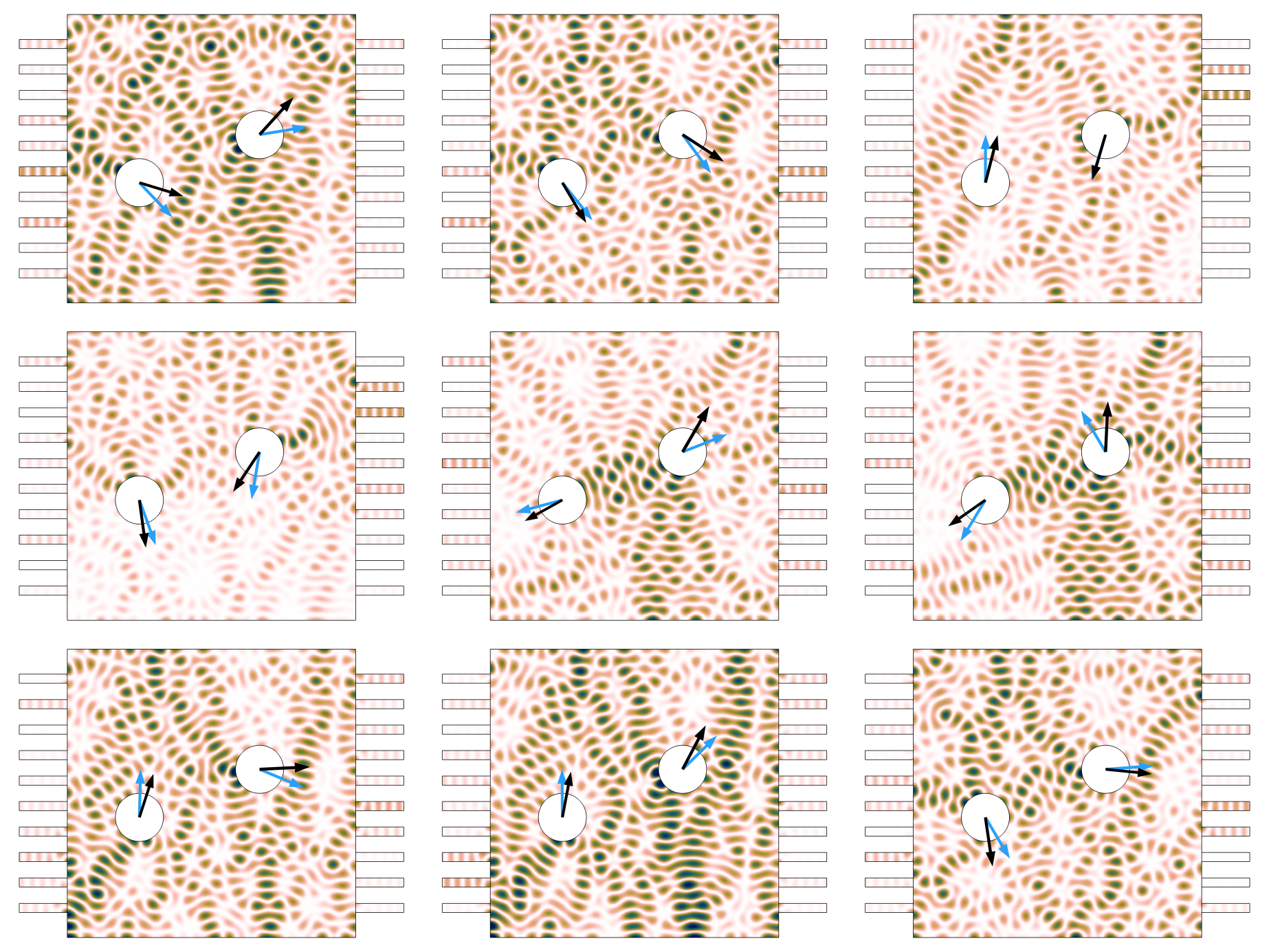}
    \caption{Further examples of bi-objective optimization for which $\braket* [1] {\bold{Q}_{\delta \bold{r}_1}}=\braket* [1] {\bold{Q}_{\delta \bold{r}_2}}$. As in \cref{fig:subfigs}b, blue arrows represent the objectives and black arrows represent the realized Pareto-optimal net equal-projection forces on the scatterers.}
    \label{fig:supp-biobj}
\end{figure}

\begin{figure}[ht]
    \centering
    \includegraphics[width=\linewidth]{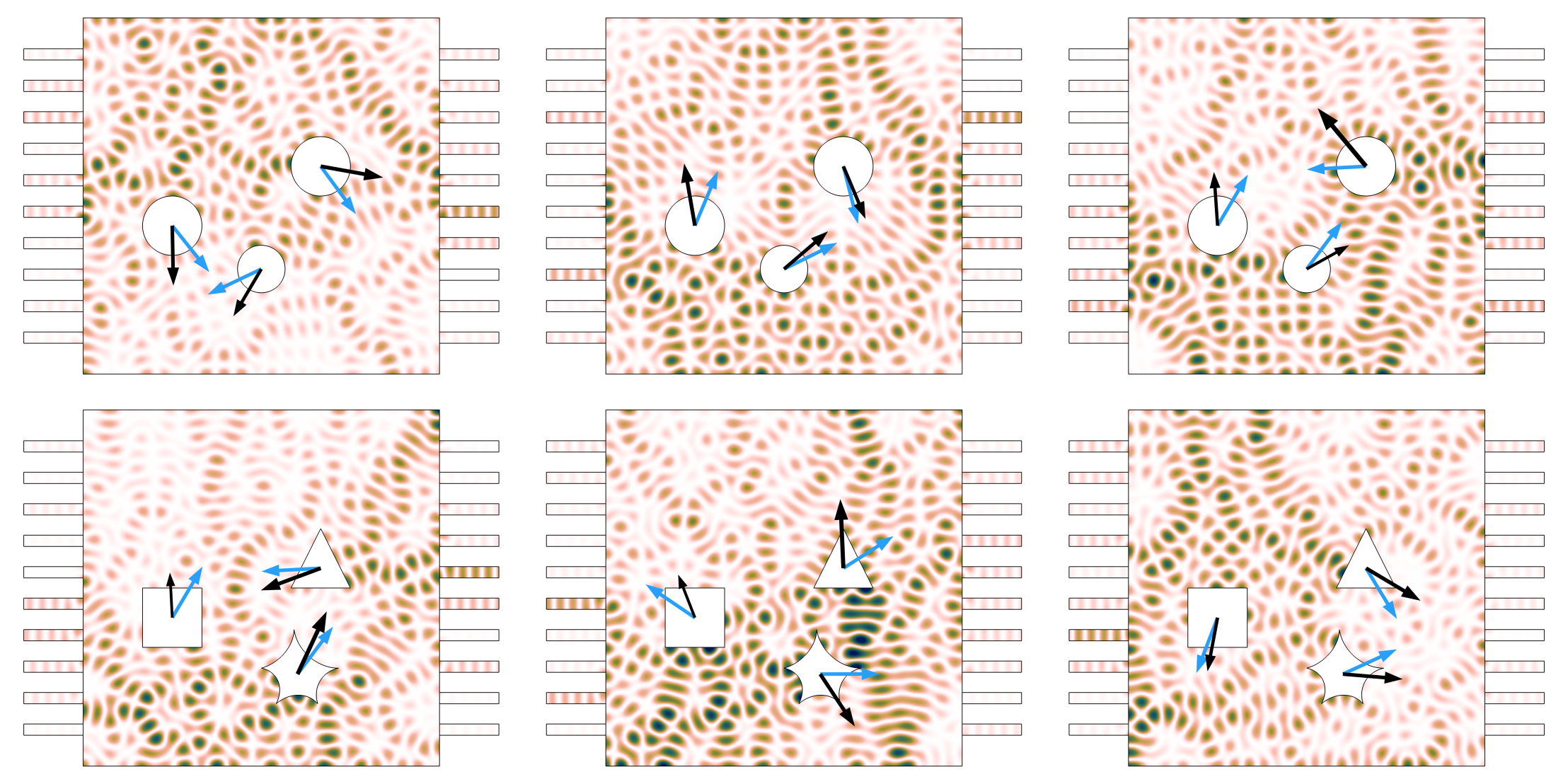}
    \caption{Examples of tri-objective optimization, for both circular and non circular objects.}
    \label{fig:supp-triobj}
\end{figure}

\subsection{Details on the cancellation of force components in the Fig.~2c of the main text.}

In Fig.~2c, the reported force directions are obtained from the Brillouin stress-tensor evaluation (and, in the electromagnetic counterpart, this could be evaluated via the Maxwell stress-tensor). To make this transparent, we compute the stress-tensor traction components along the entire particles circumferences and plot in Fig.~\ref{fig:integrals} the corresponding $t_x(\theta)$ $t_y(\theta)$ such that
\begin{equation}
    \overline{F}_x = \oint t_x(\theta) \diff{s} \quad \text{and} \quad  \overline{F}_y = \oint t_y(\theta) \diff{s},
\end{equation}
where $\diff{s}=R \diff{\theta}$ is the infinitesimal arc-length element along the object's boundary (circular objects of radius $R=1$ cm here). These data demonstrate that the azimuthal contributions to $\overline{F}_x$ on objects \protect\circled{1} and \protect\circled{2} cancel upon integration over $\theta$ (almost zero net $\overline{F}_x$), whereas $\overline{F}_y$ exhibits a nonzero average on object \protect\circled{1} (and almost zero on \protect\circled{2}), yielding a net force aligned with the vertical direction as stated in the main text.
\begin{figure}
    \centering
    \includegraphics[width=0.9\linewidth]{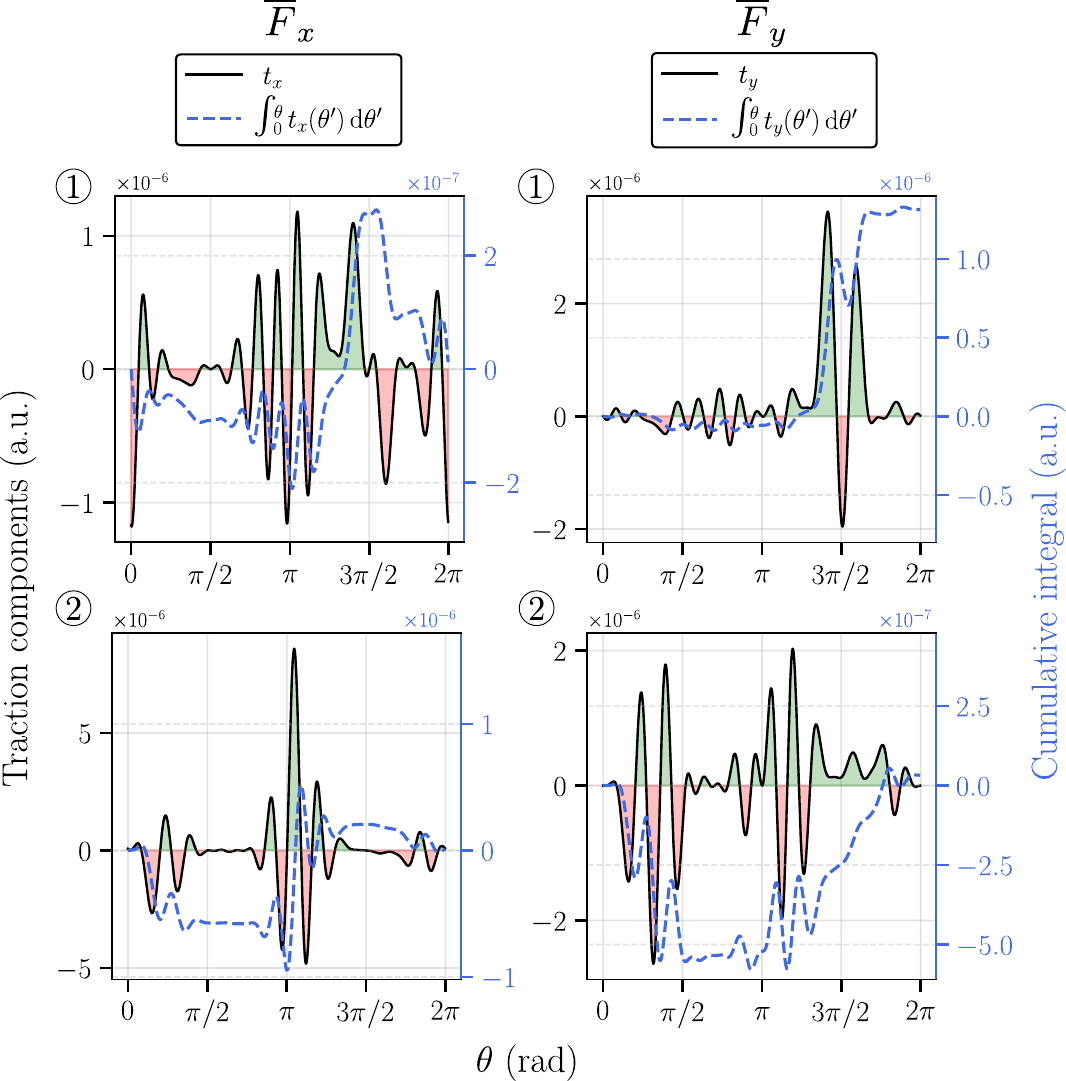}
    \caption{Local Brillouin stress-tensor traction on the particles of Fig.~2c. For each particle (top: object \protect\circled{1}; bottom: object \protect\circled{2}), the traction components along the boundary are shown as a function of polar angle, $t_x(\theta)$ (left) and $t_y(\theta)$ (right) (black). The red/green shading indicates negative/positive local contributions to the traction. The blue dashed curves show the cumulative line integral along the circumference, whose value at $\theta=2\pi$ gives the net time-averaged force component. Despite sizable local tractions, the contributions to $t_x$ cancel almost exactly for both objects, yielding $\overline{F}_x\approx 0$, whereas object \protect\circled{1} exhibits a nonzero net accumulation in $t_y$ (and object \protect\circled{2} nearly cancels), explaining the vertical force direction reported in Fig.~2c.}
    \label{fig:integrals}
\end{figure}

It is worth noticing that although the Gurobi optimization predicts a $\sim 10^{8}$ separation between ($\braket* [1]{\bold{Q}_{\delta x_1}}$,$\braket* [1]{\bold{Q}_{\delta x_2}}$, $\braket* [1]{\bold{Q}_{\delta y_2}}$) and $\braket* [1]{\bold{Q}_{\delta y_1}}$ (indicating that the numerical optimization is extremely tight), the FEM post-processing does not exhibit the same $\sim 10^{8}$ separation in the evaluation of the traction integrals. This discrepancy is expected because both the GWS matrices used in the optimization and the traction integrals evaluated in the simulation are assembled from finite-element fields: the near-cancellations of the desired forces are therefore limited by mesh discretization and numerical quadrature errors. Even with these errors, the FEM results still show that the optimized vertical force on object \protect\circled{1} is $\sim 10^{2}$ larger than the residual force components (the horizontal force on object \protect\circled{1}, the horizontal force on object \protect\circled{2}, and the vertical force on object \protect\circled{2}), which is sufficient to conclude that the response is effectively constrained to the $y$-direction on object \protect\circled{1}.
\endgroup
\end{document}